\documentclass[11pt,preprint]{aastex}

\newcommand{\kms}{\,km~s$^{-1}$}

\newcommand{\Msun}{\mbox{\,$M_{\odot}$}}

\def\spose#1{\hbox to 0pt{#1\hss}}
\def\simlt{\mathrel{\spose{\lower 3pt\hbox{$\mathchar"218$}}
     \raise 2.0pt\hbox{$\mathchar"13C$}}}
\def\simgt{\mathrel{\spose{\lower 3pt\hbox{$\mathchar"218$}}
     \raise 2.0pt\hbox{$\mathchar"13E$}}}
\font\smcap=cmcsc10
\def\caii{Ca\,{\smcap ii}}

\slugcomment{Accepted to AJ}

\shorttitle{NGC~205}
\shortauthors{Geha~et~al.}

\begin{document}


\title{Local Group Dwarf Elliptical Galaxies:  I.\ Mapping the Dynamics of
    NGC~205 Beyond the Tidal Radius}


\author{M.\ Geha\altaffilmark{1}}
\affil{The Observatories of the Carnegie Institute of Washington,
    813 Santa Barbara Street, Pasadena, CA~91101.}
\altaffiltext{1}{Hubble Fellow.}
\email{mgeha@ociw.edu}

\author{P.\ Guhathakurta}
\affil{UCO/Lick Observatory, University of California,
    Santa Cruz, 1156 High Street, Santa Cruz, CA~95064.}
\email{raja@ucolick.org}

\author{R.\ M.\ Rich}
\affil{Department of Physics and Astronomy, UCLA, Box 951547, Los Angeles,
    CA~90095.}
\email{rmr@astro.ucla.edu}

\author{M.\ C.\ Cooper} \affil{Department of Astronomy, University of California
    at Berkeley, 601 Campbell Hall, Berkeley, CA 94720-3411.}
\email{cooper@astron.berkeley.edu}


\begin{abstract}
\renewcommand{\thefootnote}{\fnsymbol{footnote}}

NGC~205 is the nearest example of a dwarf elliptical (dE) galaxy and
the prototype of this enigmatic galaxy class.  Photometric evidence
suggests that NGC~205, a close satellite of the M31 galaxy, is tidally
interacting with its parent galaxy.  We present stellar radial
velocity measurements out to a projected radius of $20'$ (5~kpc) in
NGC~205 based on Keck/DEIMOS multislit spectroscopic observations of
725~individual red giant branch stars.  Our kinematic measurements
extend from the center out to six times the effective radius of
NGC~205, well past the expected tidal radius.  The contamination in
our kinematic sample from M31 field stars is estimated to be a few
percent based on maximum likelihood fits to the distribution of stars
in position-velocity space.  We measure a maximum major-axis rotation
speed for the body of NGC~205 of $11\pm5$\,\kms\ and note that this is
based on observing a definite turnover in the rotation curve; this is
the first dE galaxy in which the maximum rotation velocity has been
measured.  Combined with the velocity dispersion, we conclude that
NGC~205 is supported by a combination of rotation and anisotropic
velocity dispersion.  At a major-axis distance of $4.5'$~(1\,kpc), the
velocity profile of NGC~205 turns over; stars beyond this radius are
moving counter to the rotation of the inner part of the galaxy.  The
turnover radius is coincident with the onset of isophotal twisting and
the estimated tidal radius, suggesting that the outer kinematics of
NGC~205 are dominated by gravitational interactions with the nearby
M31 galaxy.  The motion of stars beyond a radius of $\sim 4.5'$
implies that NGC~205 is in a prograde encounter with its parent galaxy
M31.

\end{abstract}




\section{Introduction}\label{intro_sec}
\renewcommand{\thefootnote}{\fnsymbol{footnote}}



NGC~205 is a low-luminosity, early-type, dwarf galaxy in the Local Group
and is one of the brightest nearby examples of a dwarf elliptical (dE)
galaxy.  A close satellite of the large spiral galaxy Andromeda (M31),
NGC~205 has a projected separation to M31 of merely $40'$ or 8\,kpc
(Fig.~\ref{fig_masks}); the physical separation between NGC~205 and
M31 is estimated to be $\sim 40$\,kpc \citep{dem03,mcc05}.  This
separation distance is intermediate between that of the Milky Way and
its dwarf satellite galaxies Sagittarius \citep[24\,kpc;][]{iba95} and
the Large Magellanic Cloud \citep[50\,kpc;][]{fre01}.  Observational
evidence such as isophotal twisting at large radii
\citep*{hod73,ken87,cho02}, recent star formation \citep{pel93,cap99},
and a steadily increasing velocity dispersion with radius
\citep*{ben91,sim02} suggests that NGC~205 is tidally
interacting with its parent galaxy.  However, inside the effective
radius, $r_{\rm eff} = 2.5' (0.6$\,kpc), NGC~205 exhibits regular
surface brightness and kinematic profiles typical of a normal dE
galaxy.  NGC~205 therefore offers a unique opportunity to study both
the internal dynamics of a nearby dE galaxy while providing a detailed
view of a satellite undergoing disruption.

A crucial question for current galaxy formation theories is to what
extent the accretion of dwarf satellite galaxies contributes to the
growth of massive galactic halos \citep{bul00,fon05}.  The M31 galaxy
and its associated dwarf satellite system are an excellent laboratory
to study this question in detail: M31 is the nearest massive galaxy
external to the Milky Way in which current observing techniques are
able to resolve individual stars.  \citet{iba01} and \citet{fer02}
first detected evidence for tidal streams in the halo of M31 based on
wide-area star-count maps.  \citet{guh04} and \citet{iba04} provided
confirmation that the stars in these streams are dynamically related
and have a small enough velocity dispersion to have feasibly been
stripped from a dwarf-sized galaxy.  \citet{mey01} have suggested that
the globular cluster G1, one of the brightest globular clusters
belonging to M31, is the remnant core of a nucleated dwarf galaxy
based on its internal dynamics and stellar populations (however see
\citet{rei04}).  From these observations, it is clear that dwarf
satellite disruption must contribute to the build-up of M31's halo;
detailed study of current, on-going interactions between M31 and its
satellites should help quantify what fraction of halo material is
attributed to this process.

Galaxy formation scenarios for the origin of dE galaxies lie broadly
in two very distinct categories: (1)~dEs are old, primordial objects,
and (2)~dEs have recently evolved or transformed from a progenitor
galaxy population \citep{dek86,moo98,may01,mas04}.  Dwarf elliptical
galaxies are highly clustered \citep{bin88}-- the majority of dEs in
the local Universe are found in dense galaxy clusters.  Kinematic
studies of dEs in the Virgo and Fornax Clusters suggest a large range
in their kinematic properties, from rotationally-supported galaxies to
dEs with no detectable major axis rotation \citep{geh03,der01,van04}.
These results, in particular the abundance of dE galaxies with a
complete lack of rotation, have so far been difficult to explain in
the context of current formation models.  The three bright Local Group
dE satellites NGC~205, NGC~147, and NGC~185 provide an excellent
opportunity to study dE kinematics in greater detail and to larger
radii than their cluster counterparts.  In this paper, we present
kinematics for the brightest of the Local Group dEs, NGC~205.

The absolute magnitude of NGC~205 is $M_V = -16.5$.  It has a nearly
exponential surface brightness profile \citep{cho02}, characteristic
of dE galaxies.  As a result, NGC~205 has a far more diffuse
appearance than the other bright M31 satellite galaxy, M32, which has
a similar absolute magnitude but is classified as a compact
elliptical.  Near the center of this galaxy, inside a radius of $1'$
(0.24\,kpc), NGC~205 contains dust \citep{haa98}, atomic and molecular
gas \citep{wel98,you97}, and a number of OB associations implying a
recent episode of star formation in the past 100 Myr
\citep{pel93,lee96}.  Beyond this radius, NGC~205 is gas and dust-free
and is composed of an intermediate stellar population \citep{dem03},
typical for similarly-sized dEs in the Virgo cluster \citep{geh03}.

Early dynamical studies suggested that NGC~205 had negligible internal
rotation velocity.  Dynamical measurements by \citet{ben91} placed an
upper limit on the rotation velocity of $1.5\pm0.8$\,\kms\ at $\sim r_{\rm
eff}$. Combined with observations of two other Local Group dEs, this
established the paradigm that dE galaxies are supported primarily by
anisotropic velocity dispersions.  However, \citet{sim02} more
recently measured the major-axis rotation of NGC~205 out to twice the
effective radius, measuring a rotation speed $13\pm 2\,$\kms.  The 
central velocity dispersion of NGC~205 has a value of $16\pm4\,$\kms\,
\citep{pet93,car90}, increasing with radius to 42\,\kms\ at $2'$
\citep{ben91} and 50\,\kms\ at $5'$ \citep{sim02}.  Assuming an isotropic
spheroid model, \citet{ben91} inferred a mass-to-light ratio of
$M/L\sim7$ for NGC~205, similar to the mass-to-light ratios inferred
for a sample of Virgo and Fornax Cluster dE galaxies based on more
sophisticated dynamical modeling \citep{geh02,der01}.  These ratios
are consistent with that of a modestly old stellar population with no
dark matter.  This lack of dark matter in the inner region is in
stark contrast to the much larger mass-to-light ratios inferred for
fainter dwarf spheroidal (dSph) galaxies in the Local Group, such as
the Draco and Ursa Minor dSph, which appear to be
dark-matter-dominated even in their central region \citep{kle03}.

Integrated-light spectroscopy cannot probe the kinematics of dE
galaxies much beyond the effective radius due to their characteristic
low surface brightness ($\mu_{V,\rm eff} = 23$ mag arcsec$^{-1}$).
Unlike dwarf irregular galaxies, dEs are devoid of gas at large radii
so their outer kinematics cannot be studied via HI gas dynamics.
Other kinematical probes exist at large radius, such as globular
clusters \citep{bea05} and planetary nebulae \citep{cor05}, but their
small numbers make these poor kinematic tracers.  The proximity of
NGC~205 presents a unique opportunity to study the dynamics of a dE
galaxy out to much larger radii than possible in more distant systems:
via spectroscopy of individual red giant branch (RGB) stars.

In this paper, we present accurate radial velocities for 725~RGB
stars in the Local Group dE galaxy NGC~205 measured via Keck/DEIMOS
multi-slit spectroscopy.  We present dynamics of this galaxy well
beyond the inferred tidal radius, twice the radial extent of previous
measurements.  This paper is organized as follows: in
\S\,\ref{sec_data} we discuss target selection for our DEIMOS
slitmasks, the observing procedure and data reduction.  In
\S\,\ref{sec_dyn}, a detailed kinematic analysis of NGC~205 is
presented along with a maximum-likelihood estimate of the kinematic
contamination from stars in M31.  Finally, in \S\,\ref{sec_orbit},
we discuss the implications of our kinematic observations for the tidal
interaction between NGC~205 and M31.

Throughout this paper we adopt a distance modulus to NGC~205 of $(m
-M)_0 = 24.58 \pm 0.07$, i.e.,~a distance of $824\pm 27$\,kpc, as
determined by \citet{mcc05} via the tip of the RGB
method; this places NGC~205 40\,kpc further than its parent galaxy M31.

\section{Data}\label{sec_data}

\subsection{Target Selection}\label{subsec_targets}

Stars were selected for spectroscopy according to their probability of
being a RGB star in NGC~205.  Stars were selected
based on Canada-France-Hawaii Telescope CFH12K mosaic imaging in the $R$
and $I$ bands kindly provided by \citet*[][private
communication]{dem03}.  The CCD mosaic covers a $42'\times28'$ region
centered on NGC~205 with $0.206''$ pixels.  The images were obtained in
sub-arcsecond seeing conditions.  Stellar photometry from these images
extends 2~mag
below the tip of the RGB.  The color-magnitude diagram for this region
is shown in Figure~\ref{fig_cmd}.  A broad RGB is seen extending to
very red colors ($R-I > 3.0$); the reddest of these stars are likely to be
metal-rich/reddened M31 disk RGB contaminants \citep{dem03}.  The vertical
ridge near $(R-I) = 0.2$ is due to foreground Milky Way stars.  The
spatial distribution of stars is shown in Figure~\ref{fig_imvel}.  The
gaps between CFH12K CCD chips are $\sim7''$ wide.  The apparent hole in
stellar density map near the center of NGC~205 is due to crowding; the
stellar density increase in the lower left corner is due to the
presence of field RGB stars belonging to M31.

The absolute magnitude of the RGB is $M_I\sim -4$.  At the assumed
distance of NGC~205, the apparent magnitude of this stellar population
is $I \sim 21$.  Since metallicity variations within NGC~205 will
cause a spread in the colors of its RGB stars, we select spectroscopic
targets based primarily on apparent magnitude.  Highest priority in
the spectroscopic target list was assigned to stars between $20.5 \le
I \le 21.0$.  Second priority was given to stars between $20.0 \le I <
20.5$ and $21.0 < I \le 21.5$; lowest priority was assigned to objects
$I > 21.5$.  To minimize Galactic foreground contamination, targets
were required to have $(R-I) > 0.2$.  Stars with photometric errors
larger than $\sigma_{I} > 0.1$ were rejected, as were stars having
neighbors of equal or greater brightness within a radius $r< 4''$.  The
photometric criteria used for spectroscopic target selection are shown
in Figure~\ref{fig_cmd}.

\subsection{DEIMOS Multi-Slit Mask Design}\label{ssec_maskdesign}

Four multislit masks were observed on the nights of 2003 September 30 and
October 1 with the Keck~II 10-m telescope and the DEIMOS
multi-object spectrograph \citep{fab03}.  Three of the four masks
(N205-1, N205-2, N205-3) were designed to be observed in a
conventional mode, while a fourth mask was designed to have multiple
tiers of spectra as described in \S\,\ref{sssec_multitier}.  Science
exposures for all four masks were $3\times1200$s per mask; the average
seeing FWHM during the spectroscopic observations was $0.7''$.

\begin{figure}
\plotone{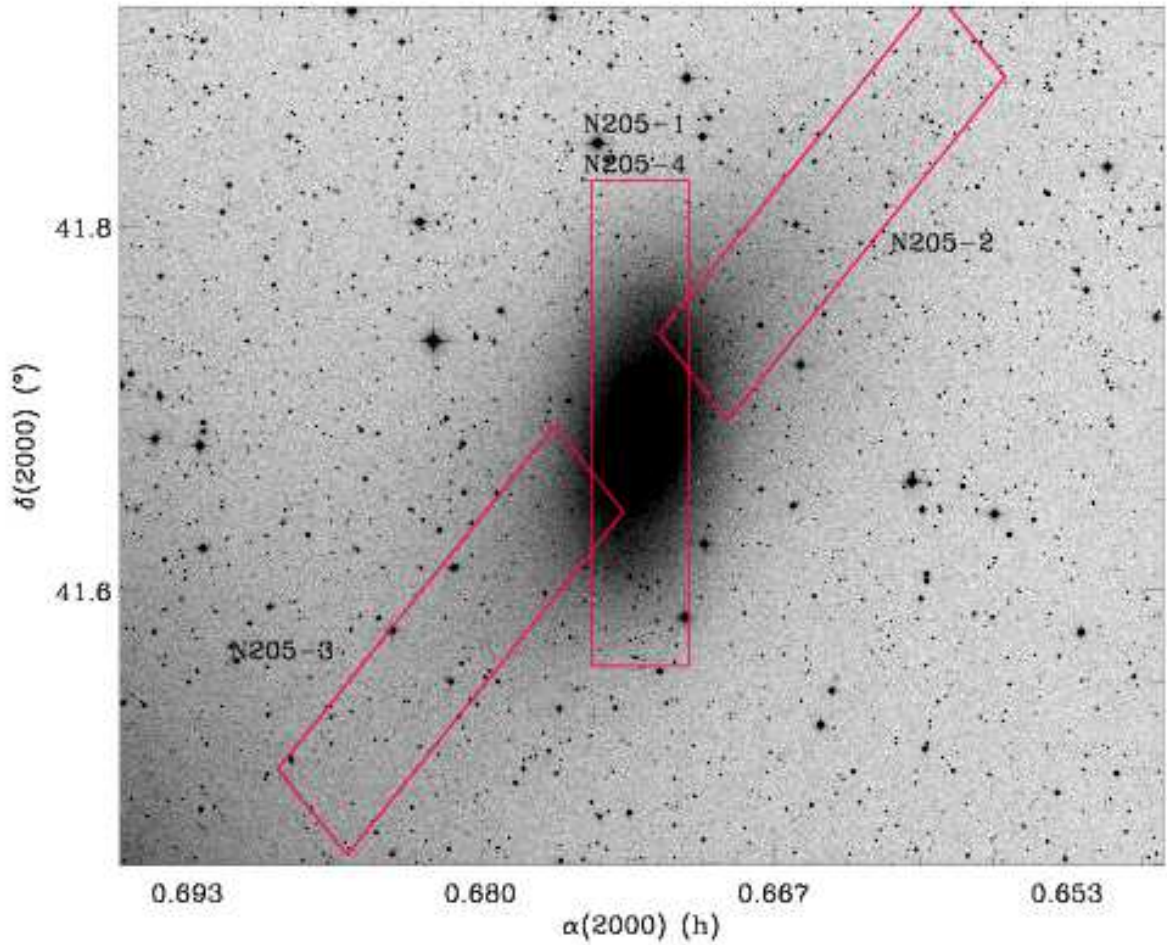}
\caption{Digitized Palomar Sky Survey image of NGC~205 showing the
  placement of the four DEIMOS slitmasks (N205-1 through N205-4).  The
  image is $42'\times 28'$; North is up, East is to the left.  The
  disk of M31 can be seen in the bottom left (South-East) corner of
  the image.
\label{fig_masks}}
\end{figure}

\begin{figure}
\plotone{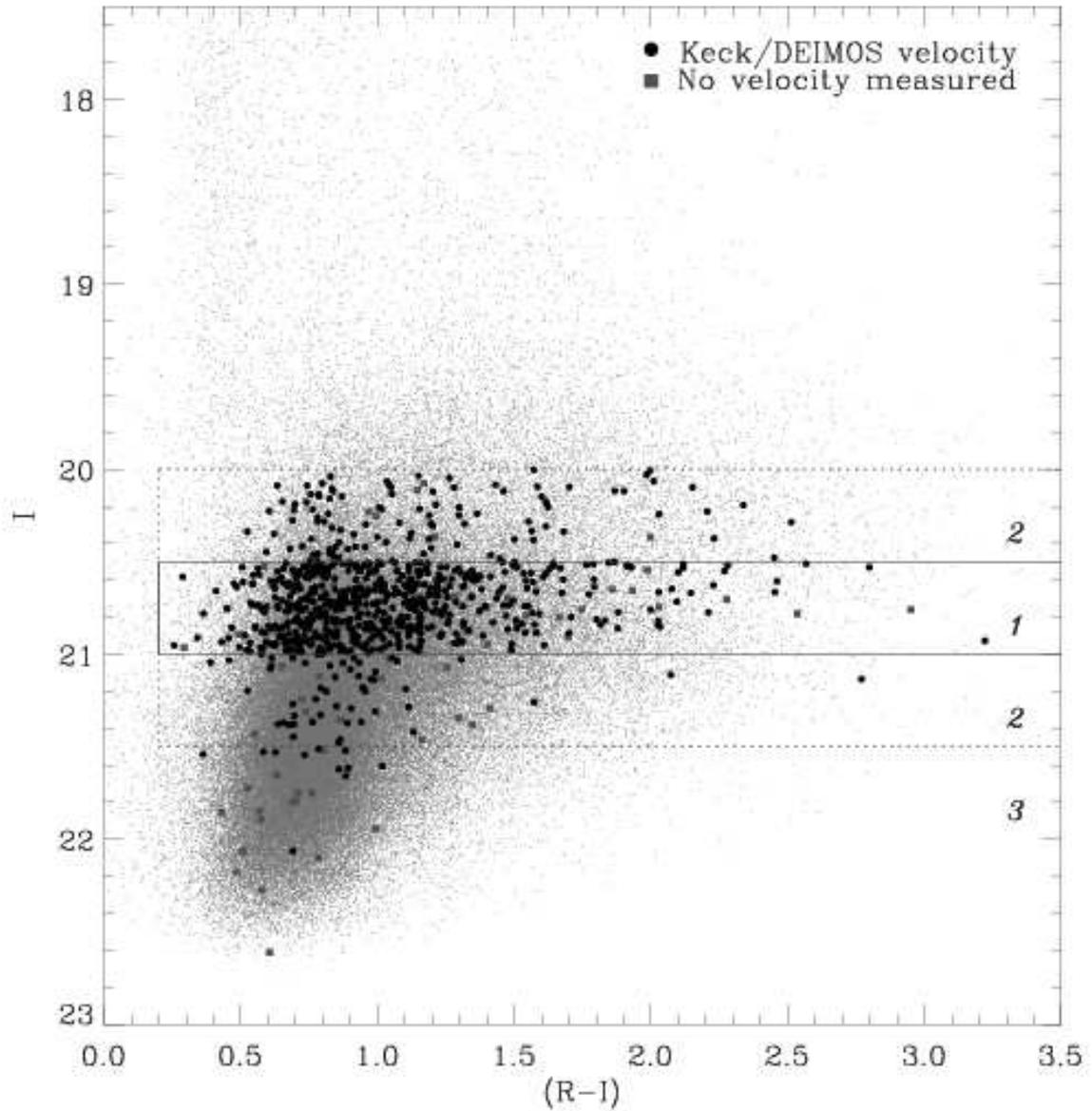}
\caption{Color-magnitude diagram based on \citet{dem03} $R$- and
  $I$-band CHFT12K photometry in a $42'\times28'$ region centered on
  NGC~205.  Large symbols indicate objects targeted for
  DEIMOS spectroscopy in this study: black circles are targets with
  measured velocities, grey squares are targets for which no velocity
  could be measured.  The numbered boxes indicate the location of our
  primary (1), secondary (2) and tertiary (3) spectroscopic
  priorities.
\label{fig_cmd}}
\end{figure}

In Figure~\ref{fig_masks}, the placement of the four DEIMOS slit-masks
is shown relative to a Digitized Palomar Optical Sky Survey image of
NGC~205.  Each DEIMOS multislit mask covers a rectangular area of
$\approx16'\times4'$.  Two slit-masks (N205-1 and N205-4) were
positioned on the center on NGC~205 with the long axis of the mask
along the major axis of the galaxy.  The centers of masks N205-2 and
N205-3 were placed $10'$ to the South-East and North-West of NGC~205,
respectively.  The long axis of these two masks were roughly placed
along the direction of tidal distortion.  The direction of tidal
distortion was determined from surface brightness ellipse fitting by
\citet{cho02} based on wide-field $B$-band photometry.  The locus of
semi-major axes of the ellipse fitting forms a gentle 'S' curve which
we plot in Figure~\ref{fig_masks} and refer to as the major-axis
of NGC~205 (see \S\,\ref{sec_dyn}).   Slitlets in each mask have the
same the position angle as the overall mask---i.e.~we did not use
tilted slits.  In Table~2, we list the observing details for each
mask.

\subsubsection{Conventional Masks}\label{sssec_conventional}

The three conventional DEIMOS slit-masks (N205-1, N205-2, N205-3) were
designed to be observed with the 1200~line~mm$^{-1}$ grating covering
a wavelength region $6400-9100\mbox{\AA}$.  The spectral dispersion of
this setup is $0.33\mbox{\AA}$, and the resulting spectral resolution,
taking into account the anamorphic distortion factor of 0.706, is
$1.37\mbox{\AA}$ (FWHM).  This wavelength range includes the \caii\
triplet lines which are expected to be strong in absorption for our
targeted RGB stars.  Slitlets were $0.7''$ wide to match the typical
seeing conditions while maintaining good wavelength resolution.  The
spatial scale is $0.12''$~per pixel and the spectral dispersion of
$0.33\mbox{\AA}$ per pixel.  To allow adequate sky subtraction, the
minimum slit length was $5''$; the minimum spatial separation between
slit ends was $0.4''$ (three pixels).

Using the above input parameters and the target list described in
\S\,\ref{subsec_targets}, slitmasks were created using the DEIMOS {\tt
dsimulator}\footnote{Available at
http://www.ucolick.org/$\sim$phillips/deimos\_ref/masks.html} slitmask
design software.  For each slitmask the software starts with the
highest priority input targets and automatically fills in the mask
area to the extent possible and filling in the remaining space on the
slitmask with lower priority targets.  An average of nearly 200
slitlets were placed on each mask (see Table~2).  A handful of targets were
deliberately included on multiple masks.  These overlapping targets
are used to quantify measurement errors in \S\,\ref{subsec_rvel}.

\subsubsection{Multi-Tier Masks}\label{sssec_multitier}

The DEIMOS slitmask N205-4 was designed for use with a blocking filter
centered on the \caii\ triplet region.  The blocking filter allows
multiple tiers of slitlets to placed on a single mask increasing the
observing efficiency by a factor of 2--3 over conventional DEIMOS
masks.  The \caii\ blocking filter has a central wavelength of
$8550\mbox{\AA}$ and is $350\mbox{\AA}$ wide.  In combination with the
831~line~mm$^{-1}$ grating, the spectral dispersion of this
setup is $0.47\mbox{\AA}$, and the resulting spectral resolution is
$1.96\mbox{\AA}$ (FWHM).  Two and a half tiers of spectra are
possible without significant overlap in the wavelength region $8450 -
8850\mbox{\AA}$.  Three full tiers of spectra are not possible because
the DEIMOS field-of-view is narrower in the middle as compared to the
ends.  The width and typical length of the slitlets are the same as
the conventional masks above.  This mask was designed by running {\tt
dsimulator} three times over the mask.  In each pass, slitlets were
placed on targets in a row running along the spatial axis of the mask
to maximize the number of slitlets.  For each pass of {\tt
dsimulator}, the region of the CCD detector plane occupied by the spectra
of all the slitlets was calculated, the spectroscopic targets in this region
excluded, and the resulting target list fed into the next pass.  A total
of 332~slitlets were placed on the N205-4 mask.  The tier pattern of
slitlets can be seen in the spatial distribution of targets in
Figure~\ref{fig_imvel}.

\subsection{Data Reduction}\label{subsec_redux}

Spectra from the three conventional DEIMOS multi-slit masks were
reduced using the {\tt spec2d} software pipeline (version~1.1.4)
developed by the DEEP2 team at the University of California-Berkeley
for that survey.  A detailed description of the reductions can be
found in \citet{coo06}.  Briefly, the flat-field exposures are used to
rectify the curved raw spectra into rectangular arrays by applying
small shifts and interpolating in the spatial direction.  A
one-dimensional slit function correction and two-dimensional
flat-field and fringing correction are applied to each slitlet.  Using
the DEIMOS optical model as a starting point, a two-dimensional
wavelength solution is determined from the arc lamp exposures with
residuals of order $\rm0.01\mbox{\AA}$.  Each slitlet is then
sky-subtracted exposure by exposure using a B-spline model for the
sky.  The individual exposures of the slitlet are averaged with
cosmic-ray rejection and inverse-variance weighting.  Finally one
dimensional spectra are extracted for all science targets using the
optimal scheme of \citet{hor86} and are rebinned into logarithmic
wavelength bins with 15\,\kms\ per pixel.  Sample one-dimensional
spectra are shown in Figure~\ref{fig_spec}.

Spectra from the multi-tier mask NGC~205-4 were reduced using a
combination of IRAF multi- and long-slit tasks similar to the method
described in \citet{geh02}.  The flat-field exposure for this mask was
used to trace the ends of each slitlet.  The APALL task was used in
``strip'' mode to extract and rectify two-dimensional rectangular
strips for each slitlet; a similar extraction was applied to the arc
lamp calibration and science frames. Data reduction proceeded on these
rectified strips.  Each strip was divided by its corresponding
normalized flat-field image.  Individual science exposures were
cleaned of cosmic rays and combined.  A wavelength solution was
determined for each slitlet from the combined Kr/Ar/Ne/Xe arc lamp
spectrum and was applied to the data.  One-dimensional spectra were
then extracted from each strip and rebinned into logarithmic
wavelength bins with 20\,\kms\ per pixel.

\subsection{Radial Velocities}\label{subsec_rvel}

Radial velocities were measured for spectra extracted from the four
DEIMOS masks by cross-correlating the observed spectra with a series
of high signal-to-noise stellar templates originally created for the
Sloan Digital Sky Survey covering a wide range of stellar spectral
type \citep{coo06}.  These templates were rebinned to match the final
wavelength resolution of the observations of 15 and 20\,\kms\ per
pixel for the conventional and multi-tier masks, respectively.  The
science and template spectra were continuum-subtracted; the template
was then shifted and scaled to minimize the reduced-$\chi^2$.  A
heliocentric correction was applied to all the measured radial
velocities.  The fitted velocities were visually inspected and
assigned a quality code to indicate the reliability of the measured
redshift and the overall quality of the spectrum.

Radial velocities for individual stars were successfully measured for
769 of the 869 extracted spectra.  Of the 100 spectra for which we did
not measure a redshift, 51 spectra were unusable due to bad columns or
vignetting and 49 spectra had insufficient signal-to-noise ratios to
determine a redshift.  Of the spectra with measured velocities, 3 were
background galaxies and 41 were duplicate measurements.  Duplicate
measurements are used to estimate our velocity error bars as discussed
below.  The final sample consists of 725~unique stellar radial
velocity measurements.

We estimate the accuracy of our radial velocity measurements via
repeat measurements of stars across the four observed masks.
Twenty-nine duplicate stars were observed across the three
conventional mask.  Twelve stars on the multi-tier mask were
duplicated on the conventional masks.  The root-mean-square (rms)
radial velocity difference between pairs of measurements for stars on
the conventional masks is 16.3~km~s$^{-1}$.  Assuming the measurement
uncertainty is the same for each member of the pair, the radial
velocity error for an individual measurement is $\sqrt{2}$ times
smaller than the rms of the difference, or $11.5$~km~s$^{-1}$.  The
rms difference between the conventional and multi-tier mask is
21.7\,\kms; this rms difference is larger than the conventional masks due to
lower spectral resolution and the smaller available wavelength region
in the multi-tier design.  Assuming velocity errors between the
conventional and multi-tier mask add in quadrature, the radial
velocity error on individual multi-tier measurements is 18.4\,\kms.
Velocity measurements for individual stars are listed in Table~3.


\section{Results}\label{sec_dyn}

The measured velocities of individual stars allow us to probe the
dynamics of NGC~205 to much larger radius than possible via
integrated-light spectroscopy.  In Figure~\ref{fig_vhist}, we show the
distribution of radial velocities of individual RGB stars in the four
Keck/DEIMOS slitmasks.  The distribution of velocities in slitmasks
N205-1 and N205-4 is centered on the systemic velocity of NGC~205.
The systemic velocity of NGC~205 is $v_{\rm sys} = -246 \pm 1$\,\kms\
based on the median velocity of stars with semi-major axis distances
less than $5'$.  The two slitmasks off-center from NGC~205 show skewed
velocity distributions.  The mask N205-2, placed to the NW of NGC~205
on the farside from M31, is skewed toward more positive velocities
than the NGC~205 systemic velocity; mask N205-3, to the SE of NGC~205
nearer to M31, has more negative velocities relative to the NGC~205
systemic.  To illustrate this observation, we fit a Gaussian profile
to the combined distribution of velocities (Figure~\ref{fig_vhist},
bottom panel).  The fitted Gaussian has a central velocity of
$-246$\kms\, in agreement with the systemic velocity of NGC~205
measured above, and a velocity width of $42$\kms\, in agreement with
the average velocity dispersion of NGC~205 measured by \citet{sim02}.
In the four top panels of Figure~\ref{fig_vhist} we plot the scaled
Gaussian profile; the excess of stars in mask N205-2 and N205-3 is
clearly visible to the right and left, respectively of the main
velocity peak.  In all four masks, there are a small number of stars
with velocities less negative than NGC~205 due to foreground
contamination, while the tail of stars to more negative velocities are
primarily M31 halo stars whose systemic velocity is $-300$\,\kms.  We
estimate M31 contamination fractions in our sample in
\S\,\ref{subsec_m31}.

\begin{figure}
\plotone{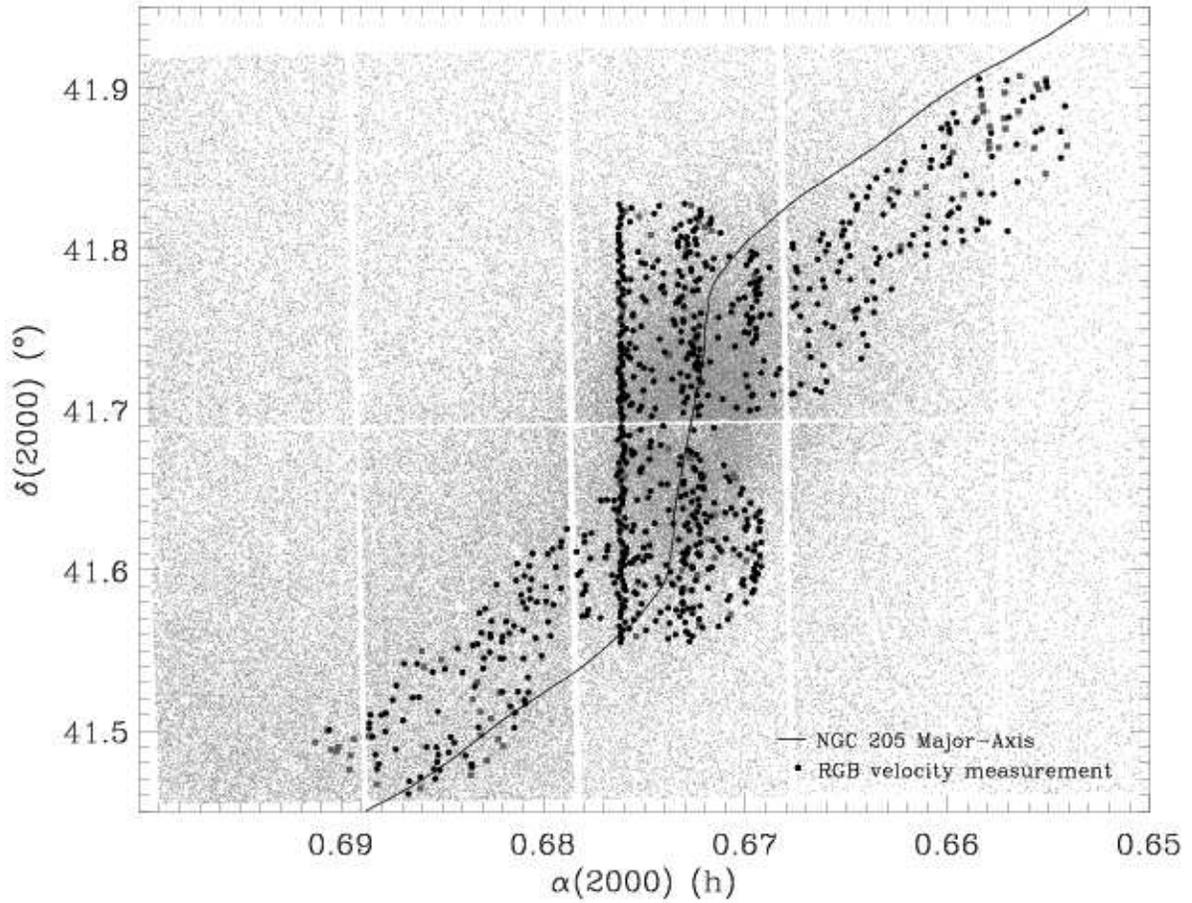}
\caption{Spatial distribution of stars in NGC~205 (small grey points)
  from \citet{dem03} photometry. Large symbols indicate objects
  targeted for DEIMOS spectroscopy; the tiered pattern of slitlets can
  be seen in the central mask.  The solid curve is the major axis of
  NGC~205 determined from \citet{cho02} photometry.
\label{fig_imvel}}
\end{figure}

\begin{figure}
\plotone{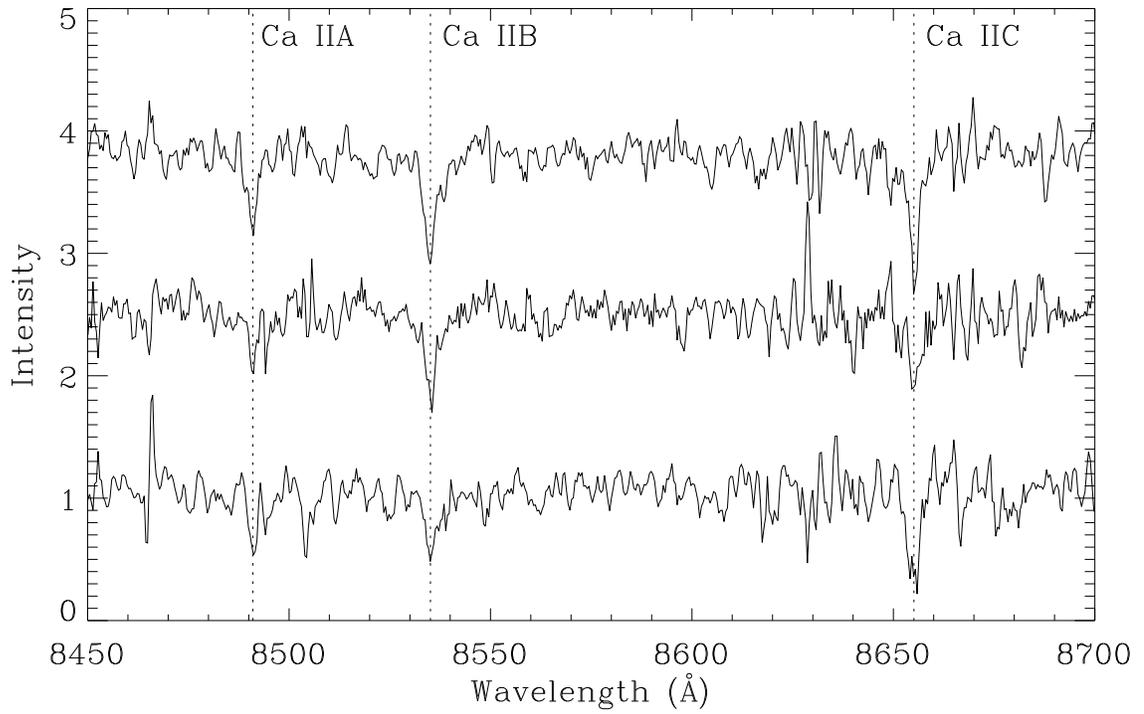}
\caption{Representative one-dimensional Keck/DEIMOS spectra of RGB
  stars in NGC~205 centered on the \caii\ region.  We compare spectra from three
  slitlets in the inner region of NGC~205.  
\label{fig_spec}}
\end{figure}

\begin{figure}
\epsscale{0.85}
\plotone{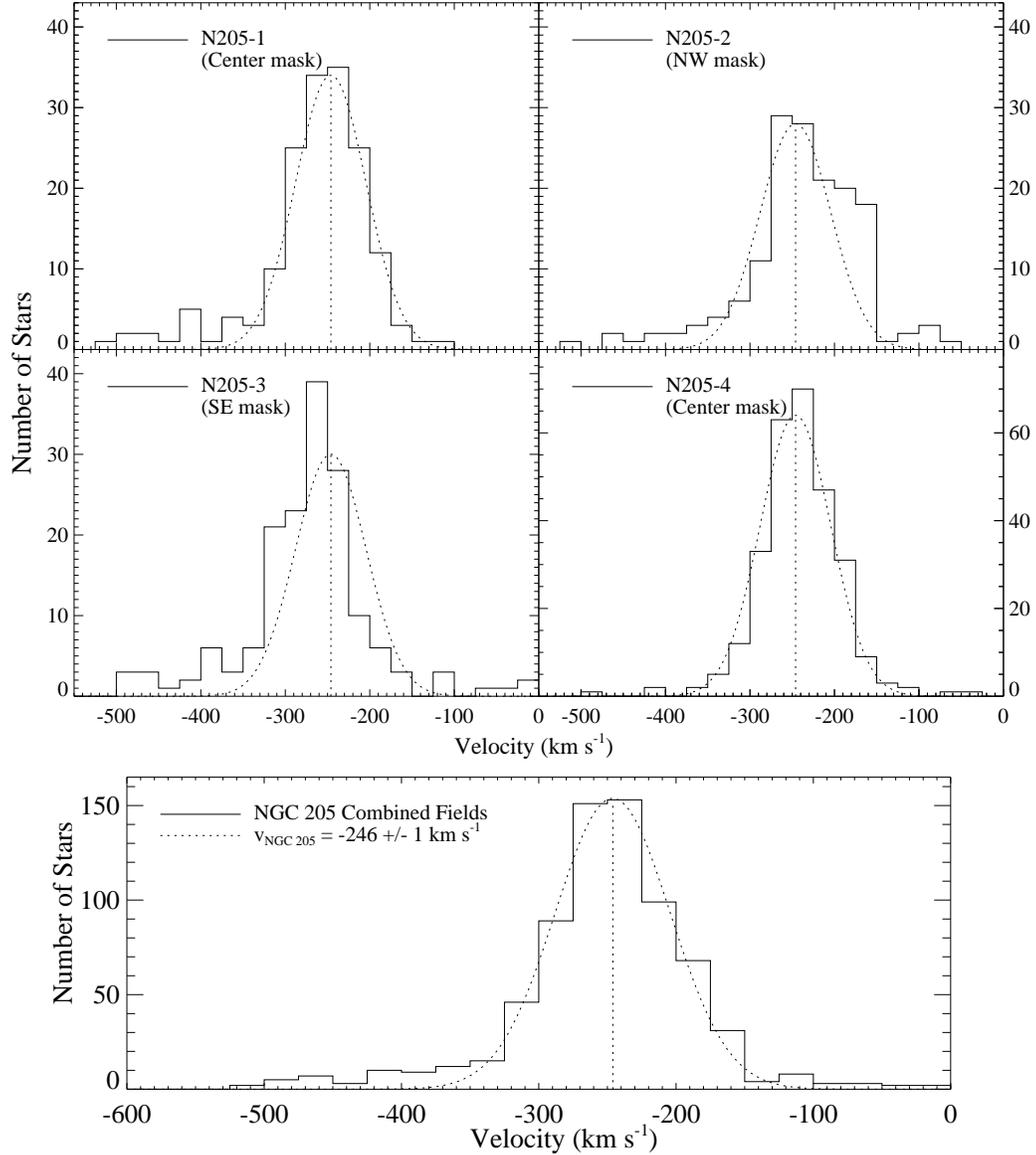}
\caption{Velocity histograms for the four observed DEIMOS masks in
  NGC~205 ({\it top four panels}) and the combined fields ({\it
  bottom\/}).  A Gaussian profile, fit to the combined velocity
  distribution, is plotted as the dotted curve in each panel.  The
  Gaussian profile has a central velocity of $v_{\rm sys} = -246$\kms\ and a
  velocity width of $42$\kms\,.  The height of the Gaussian has been
  scaled to the distribution of velocities in each individual panel.
  The masks N205-2 and N205-3, located on opposite sides of the center
    of NGC~205 along the major axis/tidal extension, display asymmetry
    in their velocity histograms towards more positive and negative
    velocities, respectively.  This is a result of internal rotation and
    tidal distortion in the dE galaxy.
\label{fig_vhist}}
\end{figure}

A semi-major axis distance is assigned to each slitlet based on its
minimum distance to the major axis of NGC~205.  The major-axis of
NGC~205 is shown in Figure~\ref{fig_imvel} and was determined from
surface brightness ellipse fitting by \citet{cho02} based on
wide-field $B$-band images of NGC~205, corrected for the underlying
light of M31.  The major-axis of NGC~205 forms a gentle 'S' curve as
plotted in Figure~\ref{fig_imvel}.  For each slitlet, we determine its
minimum distance to the major-axis and assign the semi-major axis
length (a) at the point.  Although there are alternative methods to
assign this distance (e.g.~the semi-major distance corresponding to
the nearest elliptical isophote), this method most closely resembles
integrated-light spectroscopy of more distant dEs to which we will
compare our results.

\subsection{The Velocity Profile of NGC~205}\label{vp_prof}

\begin{figure}
\epsscale{1.0}
\plotone{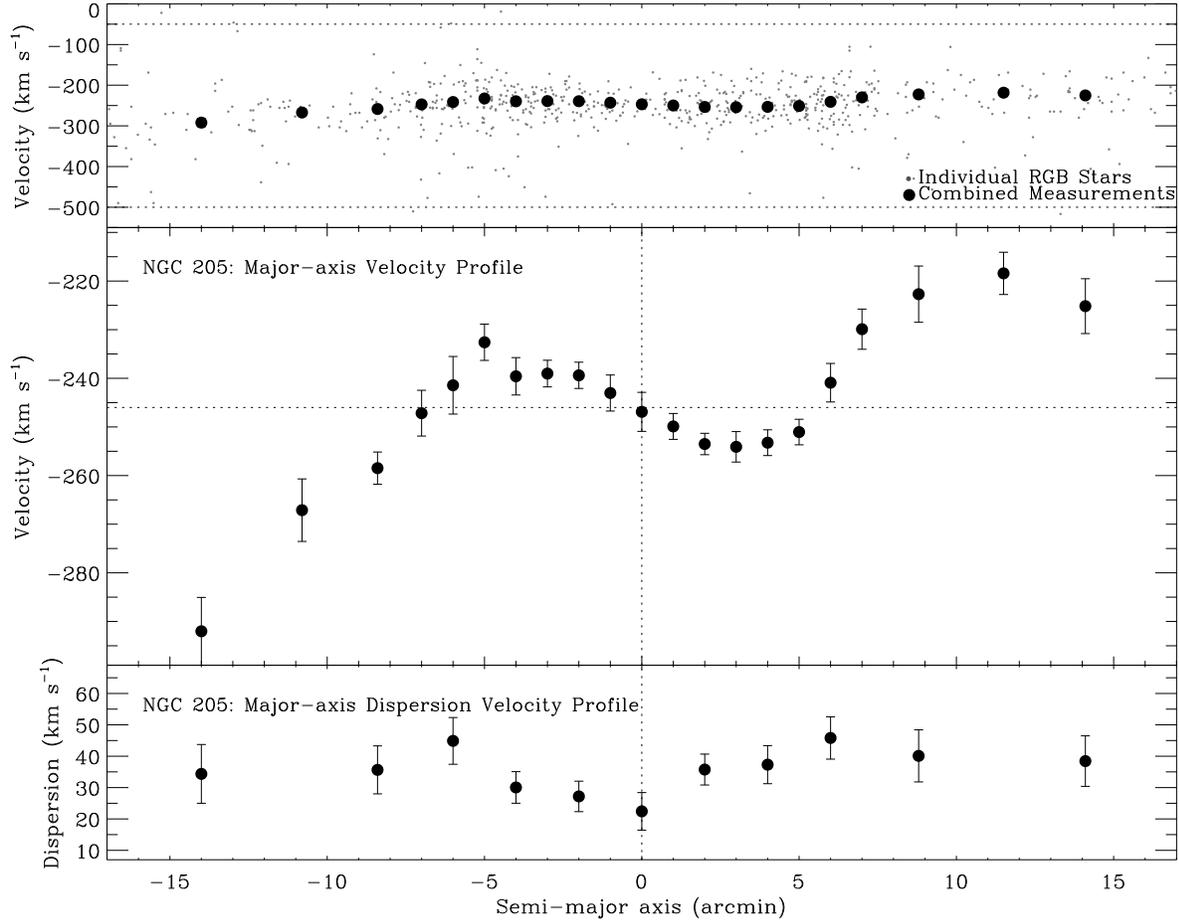}
\caption{Major-axis velocity profile for NGC~205.  ({\it Top\/}) Small
  grey symbols indicate Keck/DEIMOS velocity measurements of
  individual RGB stars, larger black symbols are the combined velocity
  measurements based on Gaussian fits to the velocity distribution in
  each radial bin.  .  Dotted lines indicate the limits inside which
  the combine measurements are determined.  ({\it Middle\/}) Combined
  velocity measurements (same as top panel) with a finer velocity
  scale.  The vertical dotted line is plotted at galaxy center; the
  horizontal dotted line is plotted at the measured systemic velocity
  of NGC~205. ({\it Bottom\/}) Velocity dispersion profile for NGC~205
  determined using a coarser binning scheme than the velocity profile.
\label{fig_vp}}
\end{figure}

In Figure~\ref{fig_vp}, we present the major-axis velocity profile for
NGC~205 determined from the combined velocity measurements of RGB
stars.  Individual stellar velocity measurements are shown in the top
panel of this figure.  To determine the ensemble major-axis velocity
profile, individual measurements were binned into minimum $1'$ radial
bins with 25 or more stars per bin.  The velocity in each bin was
determined by simultaneously fitting a double Gaussian profile to the
distribution of stars in that bin.  To account for contamination from
M31 stars, one of the two fitted Gaussian profiles had a fixed mean
and velocity width of -300 \kms\ and 150\kms, respectively,
corresponding to best estimates for M31's halo (see
\S\,\ref{subsec_m31}).  We fit for the width and mean of the second
Gaussian profile to determine both the velocity and velocity
dispersion of NGC~205 in each radial bin.  The relative height of
these two profiles is fixed based on the M31 contamination fraction
determined by maximum likelihood fitting in \S\,\ref{subsec_m31}.  We
have also run fits in which the contamination fraction is a free
parameter which does not change the resulting profile appreciably.
The resulting velocity and velocity dispersion profile as a function
of radius is shown in the bottom two panels of Figure~\ref{fig_vp}.
The velocity dispersion profile is determined using a coarser binning
scheme of 50 or more stars per radial bins.  Error bars were computed
based on number statistics in each radial bin which dominate over the
radial velocity measurement errors of individual stars.  We list the
velocity and velocity dispersion as a function of semi-major axis
distance, right ascension and declination in Table~4.

We determine the maximum major-axis rotation velocity, $v_{\rm max}$,
for NGC~205 by differencing the maximum/minimum velocity in the
upper-left quadrant and lower-right quadrant of Figure~\ref{fig_vp},
and dividing this number by two.  The maximum rotation velocity for
NGC~205 is $v_{\rm max} = 11 \pm 5$\,\kms.  The peak rotation velocity
occurs at slightly different radii on either side of the galaxy, with
an average radius of $r_{\rm max} = 4.5'~(1.1$\,kpc).  Our value is
somewhat smaller, but within the 1-sigma error bars, of the value
measured by \citet{sim02} of $13\pm2$\,\kms\ at $r=4'$ based on
integrated-light measurements.  The Simien \& Prugniel data hint at a
flattening in the velocity profile, but the integrated-light data is
noisy at these large radii.  We observe this turnover conclusively
and note that $v_{\rm max}$ is the physical, rather than a
observational, maximum rotation velocity.  Previous observations 
of dE galaxies have not reached sufficient radii to observe a
definite turnover in the rotation curve; this is the first dE galaxy
in which the {\it maximum} rotation velocity has been measured.

\begin{figure}
\epsscale{0.7}
\plotone{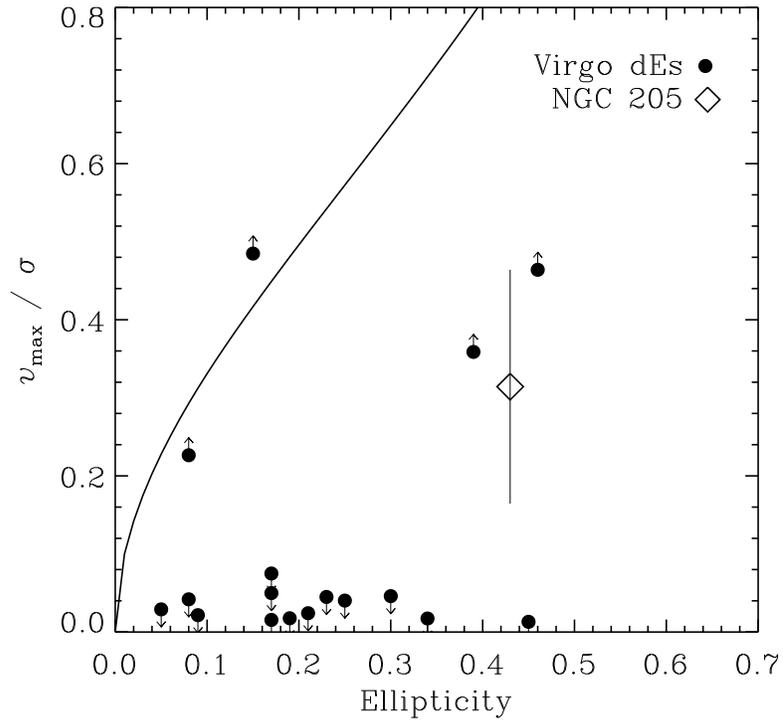}
\caption{The ratio of the rotation velocity $v_{\rm max}$ to velocity
dispersion $\sigma$ plotted versus mean isophotal ellipticity.  The
solid line is the expected relation for an oblate, isotropic galaxy
flattened by rotation.  Solid symbols indicate Virgo Cluster dEs from
\citet{geh03}, the open diamond is NGC~205.  Unlike the Virgo dEs,
NGC~205 has a mixture of both rotational and anisotropic support.
\label{fig_vsigma}}
\end{figure}

We plot the ratio of the maximum rotational velocity to the average
velocity dispersion ($v_{\rm max}/\sigma$) versus ellipticity in
Figure~\ref{fig_vsigma}.  We assume an ellipticity for NGC~205 of
$\epsilon = 0.43$ \citep{cho02}, and a velocity dispersion of $\sigma
= 35\pm5$\,\kms determined from our profile excluding data
beyond the tidal radius.  The observed ratio for NGC~205 is $v_{\rm
max}/\sigma = 0.21$.  At the ellipticity of NGC~205, the expected
ratio for an oblate, isotropic, rotationally-flattened body seen
edge-on is slightly more than unity \citep{bin87}.  NGC~205 lies
midway between a rotationally supported and anisotropic object.
Interestingly, it was early kinematic observations of NGC~205 which
established the paradigm that dE galaxies are supported by anisotropic
velocity dispersions in contrast to rotationally supported normal
ellipticals \citep{ben90}.  In Figure~\ref{fig_vsigma}, we compare
NGC~205 to a sample of similar luminosity dEs in the Virgo Cluster
taken from \citet{geh03}.  The Virgo dEs naturally fall into
``rotating'' and ``non-rotating'' categories.  NGC~205 does not fall
into either category having a mixture of both rotational and
anisotropic support.

At a radius of $r\sim4.5'$, the velocity profile of NGC~205 turns
over; stars beyond this radius rotate in a direction opposite that of
the main galaxy body.  This turnover is in the sense that stars on the
side of NGC~205 closest to M31 move with more negative velocities
(approaching the systemic velocity of M31) and stars on the farside
move with less negative velocities.  The turnover radius is coincident
with the onset of isophotal twisting as observed by \citet{cho02},
suggesting that the outer dynamics of NGC~205 have been affected by
the processes of tidal stripping.  We discuss further evidence for
tidal interactions and its significance in \S\,\ref{sec_orbit}.

\subsection{Contamination from M31 Stars and Maximum-Likelihood Analysis}\label{subsec_m31}

NGC~205 lies merely $40'$ (8\,kpc) in projection from the center of
M31 and our kinematic sample is likely contaminated with stars
associated with M31.  We consider two sources for contamination, the
M31 disk and halo.  We estimate that contamination from M31's disk is
negligible: NGC~205 lies behind M31 and its line-of-sight intersects
the M31 disk at a point 37\,kpc (7 scale lengths) along the disk from
the center of M31 (assuming an inclination angle of the M31 disk of
$12.5^{\circ}$ and a disk scale length of 5.3\,kpc \citep{wal88}).  At
this scale length, the M31 disk surface brightness is $\mu_B =
29.2$\,mag arcsec$^{-2}$.  In comparison, at the average radial
distance of our spectroscopic sample ($r = 5'$), the surface
brightness of NGC~205 is $\mu_B = 25.4$\,mag arcsec$^{-2}$.  Thus, our
kinematic sample is unlikely to contain stars from the disk of M31.

\begin{figure}
\epsscale{1.0}
\plotone{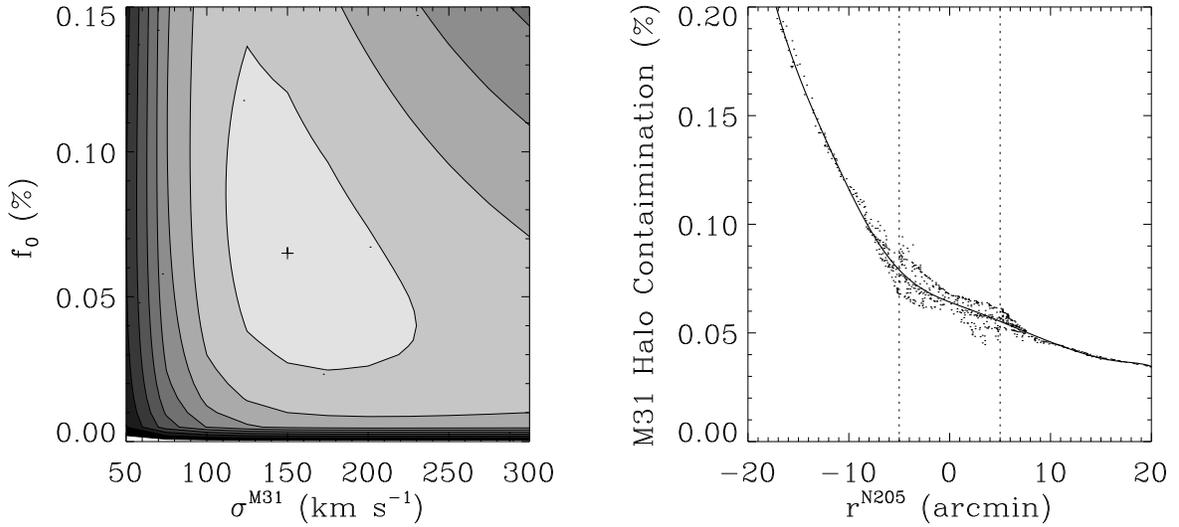}
\caption{({\it Left\/}) Likelihood contours of the contamination
fraction of M31 stars ($f_0$) versus the velocity dispersion of the
M31 halo.  The third free parameter in our model, the power-law index
of the M31 halo profile, is fixed at the best-fit value of $\alpha =
-2.25$.  The ``+'' symbol indicates the best fitting model parameters
of $f_0 = 0.065$ and $\sigma_{\rm M31} = 150$\kms.  ({\it Right\/})
M31 halo contamination fraction as a function of position in NGC~205.
The average contamination from M31 halo stars for the sample is
6.5\%.  The vertical dotted lines indicate the radius of the
turnover in the major-axis velocity profile.
\label{fig_maxlike}}
\end{figure}

We estimate the contamination fraction from stars in the M31 halo
based on maximum likelihood fitting.  We assume the M31 halo is
spherically symmetric with a power-law surface brightness profile.  We
normalize the surface brightness (SB) of the halo at the center of NGC~205,
defining the free parameter $f_0$ to be the fractional contamination
at the center of NGC~205.  The M31 halo surface brightness at the
position of a given slitlet $i$ is then:
\begin{equation}
{{\rm SB}_i}^{\rm M31} = f_0 ~ {\rm SB}^{\rm N205}_0  ~\Big{(}\frac{r^{\rm M31}_i}
       {r^{\rm M31-N205}}\Big{)}^\alpha
\end{equation}
\noindent
where the angular distance between M31 and NGC~205 is $r^{M31-N205} =
40'$, the angular distance from slitlet $i$ to the center of M31 is
$r^{\rm M31}_i$ , and the free parameter $\alpha$ is the power-law
index of the halo profile.  The surface brightness profile of NGC~205
is based on a high-order fit to the photometry of \citet{cho02} with a
central $B$-band surface brightness of $SB^{\rm N205}_0 = 19.2$~mag
arcsec$^{-2}$.  We assume a Gaussian velocity distribution for the M31
halo with a velocity dispersion of width $\sigma^{\rm M31}$ and a
systemic velocity of $v^{\rm M31}_{\rm sys} = -300$\,\kms\
\citep{guh04}.  Given a measured velocity of $v_i$ for slitlet $i$,
the probability that it was drawn from the velocity distribution
of the M31 halo is:
\begin{equation}
P_i^{\rm M31} = \frac{1}{\sqrt{2\pi} \sigma^{\rm M31}}
exp\Big{[}-\frac{1}{2}\Big{(}\frac{v_i - v^{\rm M31}_{\rm
sys}}{\sigma^{\rm M31}}\Big{)}^2\Big{]}
\end{equation}
\noindent
Similarly, the probability that a given velocity is drawn from
velocity distribution of NGC~205, $P_i^{\rm N205}$, is the same as
Eqn.~(2), however $\sigma^{\rm N205}$ and $v^{\rm N205}_{\rm sys}$ are
the measured quantities as a function of radius.  The final probability
function for each slitlet $i$ is:
\begin{equation}
P_i = C_i [{\rm SB}^{\rm N205}_i P^{\rm N205}_i + {\rm SB}^{\rm M31}_i P^{\rm
M31}_i]
\end{equation}
\noindent
The normalization constant $C_i \equiv 1/(SB^{N205}_i + SB^{M31}_i)$
is defined such that the integral over $P_i$ is equal to unity.  The
probability $P_i$ is evaluated for each slitlet and the natural
logarithm of this quantity is summed over all slits.  We do a gridded
parameter search over the free parameters $f_0$, $\sigma^{M31}$, and
$\alpha$ and maximize the function $M = \sum_i {\rm ln}(P_i)$.  As
shown in the left panel of Figure~\ref{fig_maxlike}, the best fitting
parameters which maximize the quantity $M$ are: $f_0$ = 0.065,
$\sigma_{M31} = 150$\,\kms\ and $\alpha=-2.25$.  We note that these best
fitting M31 halo parameters are in good agreement with the measured
M31 halo velocity dispersion by \citet{guh04} and an $r^{-2}$ halo
surface brightness profile.

In the right panel of Figure~\ref{fig_maxlike}, we plot the fractional
contamination in our kinematic sample from the M31 halo as a function
of distance from the center of NGC~205 using the best fit parameters
determined above.  At the center of NGC~205, the best estimate of the
contamination from M31 is 6\%; this corresponds to 1.5 out of 25 stars
per radial bin.  On the side of NGC~205 closest to M31 the
contamination is 20\%, while on the farside the contamination is less
than 3\%.  The average halo contamination fraction for our sample is
6.5\%.

\section{The Tidal Disruption of NGC~205}\label{sec_orbit}

The outer dynamics of NGC~205 place significant constraint on the
tidal interaction between NGC~205 and its parent galaxy M31.  While
the main body of NGC~205 is rotating with a maximum velocity of
11\,\kms\,, the major-axis velocity profile turns over abruptly at a
radius of $4.5'~(1$\,kpc); stars at larger radii are moving in the
opposite direction than the main galaxy body.  The turnover radius
coincides with the onset of isophotal twisting \citep{hod73,ken87} and
a downward break in the surface brightness profile \citep{cho02}.
Both features have been demonstrated by Choi~et~al.~to be hallmarks of
tidal stripping.  We make a simple estimate of the tidal radius of
NGC~205, assuming NGC~205 and M31 to be point masses.  We adopt a
physical separation distance between these two galaxies of
$r_{N205-M31} = 40$\,kpc, the mass of M31 inside this distance of
$M_{\rm M31} = 1\times10^{11}\Msun$ \citep{gee05} and the total mass
of NGC~205 $M_{\rm N205} = 2\times10^{9} \Msun$ (calculated from the
maximum rotation velocity and velocity dispersion at the turnover
radius).  The tidal radius is calculated to be $r_{\rm tidal} \sim 4'$
(1.0\,kpc).  The observed velocity turnover in NGC~205, at roughly
this distance, is therefore due to tidal stripping from M31; stars
beyond this radius are no longer bound to NGC~205.

The stellar motion beyond the tidal radius of NGC~205 suggests that it
is on a prograde encounter with its parent galaxy: the spin angular
momentum vector and the orbital angular momentum vector are
parallel.  The eventual destruction of a satellite on a
prograde orbit is expected to proceed more slowly and create less
spectacular tidal tails than one on a retrograde orbit
\citep[e.g.,~compare Figs.~1 and 2 of][]{too72}.  We calculate the
timescale for the destruction of NGC~205 based on the dynamical
friction timescale.  Using Eqn.~(7.26) in \citet{bin87}, we assume
the orbital velocity of NGC~205 is related the velocity difference
between NGC~205 and M31 ($=\sqrt{3} \times 60$\,\kms) and calculate
$t_{\rm fric} \sim 3 \times 10^9$ years.  Detailed dynamical modeling
is required to determine if features in the M31 halo, such as stellar
substructure \citep{mcc04}, or HI clouds \citep{thi04} are indeed 
associated with the orbit of NGC~205.  

\section{Summary}\label{sec_disc}

We present the stellar kinematics of NGC~205 out to large radii based
on Keck/DEIMOS multislit spectroscopic observations of 725 individual
red giant branch stars.  NGC~205 is one of the closest examples of a
dwarf elliptical galaxy and the prototype of this galaxy class.  Early
kinematic observations of NGC~205 established the paradigm that dE
galaxies are supported primarily by anisotropic velocity dispersions
\citet{ben91}.  Although kinematic study of dEs outside the Local
Group suggest that a fraction of dEs are indeed supported by
anisotropic velocity dispersion alone \citep{geh03,der01}, NGC~205
itself has significant rotation.  We measure a maximum major-axis
rotation speed for the body of NGC~205 of $11\pm5$\,\kms, implying
that this galaxy is supported by a mixture of both rotational and
anisotropic velocities.

The velocity profile of NGC~205 turns over at a major-axis distance of
$4.5'~(1$\,kpc).  This turnover is due to gravitational interaction
between NGC~205 and its parent galaxy M31.  The motion of stars beyond
the tidal radius suggest that NGC~205 is on a prograde encounter
with M31. Detailed dynamical modeling should clarify what, if any,
substructure in the M31 halo is associated with disruption of NGC~205
and provide insight into the on-going interactions between these two
galaxies.

\acknowledgments 

 We thank S.~Demers and P.~Battinelli for kindly providing their
 photometric catalogs.  We also thank P.~Choi, G.~Laughlin, D.~Kelson
 and R.~P.~van der Marel for productive and enjoyable conversations.
 M.~G.~is supported by NASA through Hubble Fellowship grant
 HF-01159.01-A awarded by the Space Telescope Science Institute, which
 is operated by the Association of Universities for Research in
 Astronomy.  P.G. acknowledges support from NSG grant AST-0307966 and
 NASA/STScI grant GO-10265.02. R.M.R. acknowledges funding by grant
 AST-0307931.




\clearpage

\begin{deluxetable}{lcccccccccc}
\tabletypesize{\scriptsize}
\tablecaption{NGC 205 at a Glance}
\tablewidth{0pt}
\tablehead{
\colhead{Name} &
\colhead{$\alpha$ (J2000)} &
\colhead{$\delta$ (J2000)} &
\colhead{Type} &
\colhead{Dist.} &
\colhead{$m_B$} &
\colhead{$M_{B,0}$}&
\colhead{$\epsilon$}&
\colhead{$\mu_{B,\rm eff}$}&
\colhead{$r_{\rm eff}$} &
\colhead{$r_{\rm tidal}$} \\
\colhead{}&
\colhead{(h$\,$:$\,$m$\,$:$\,$s)} &
\colhead{($^\circ\,$:$\,'\,$:$\,''$)} &
\colhead{}&
\colhead{(kpc)}&
\colhead{}&
\colhead{}&
\colhead{}&
\colhead{(mag arcs$^{-2}$)}&
\colhead{[$'$ (kpc)]}&
\colhead{[$'$ (kpc)]}
}
\startdata
NGC~205  &  00:40:22 & +41:41:07 & dE5 & 824 & 9.9 & $-15.0$  & 0.43 & 23.7 & 2.5 (0.6)  &  4.5 (1.1) \\
\enddata

\tablecomments{The right ascension, declination, and morphological
type of NGC~205 are take from NASA/IPAC Extragalactic Database (NED).
We adopt the distance to NGC~205 determined by \citet{mcc05}.  The
apparent magnitude, effective surface brightness, effective radius,
and ellipticity are determined from $B$-band photometry from
\citet{cho02}.  The ellipticity $\epsilon$ is the average value
measured between $0.5'$ and $5'$.  The absolute magnitude is calculated
assuming an extinction of $A_B = 0.27$ from \citet{sch98}.  The tidal
radius $r_{\rm tidal}$ is calculated based on the assumptions outlined
in \S\,\ref{sec_orbit}}.
\end{deluxetable}

\begin{deluxetable}{lccrcccc}
\tabletypesize{\scriptsize}
\tablecaption{Keck/DEIMOS Multi-Slitmask Observing Parameters}
\tablewidth{0pt}
\tablehead{
\colhead{Mask} &
\colhead{$\alpha$ (J2000)} &
\colhead{$\delta$ (J2000)} &
\colhead{PA} &
\colhead{$t_{\rm exp}$} &
\colhead{\# of slits} &
\colhead{\% useful} \\
\colhead{Name}&
\colhead{(h$\,$:$\,$m$\,$:$\,$s)} &
\colhead{($^\circ\,$:$\,'\,$:$\,''$)} &
\colhead{(deg)} &
\colhead{(sec)} &
\colhead{}&
\colhead{spectra}
}
\startdata
N205-1  & 00:40:27.4  & +41:41:30.5  &  0  & 3600 & 184 & 91\%\\
N205-2  & 00:39:50.5  & +41:48:41.8  & $-45$ & 3600 & 185 & 85\%\\
N205-3  & 00:40:53.0  & +41:40:53.0  & $-45$ & 3600 & 193 & 84\%\\
N205-4  & 00:40:27.4  & +41:41:30.5  &  0  & 3600 & 320 & 88\%\\
\enddata

\tablecomments{Right ascension, declination, position angle and total exposure
time for each Keck/DEIMOS slitmask.  The final two columns refer to
the total number of slitlets on each mask and the percentage of those
slitlets for which a redshift was measured.}
\end{deluxetable}

\begin{deluxetable}{lccccrccc}
\tabletypesize{\scriptsize}
\tablecaption{Velocity Measurements for Individual Stars in NGC 205}
\tablewidth{0pt}
\tablehead{
\colhead{Name} &
\colhead{$\alpha$ (J2000)} &
\colhead{$\delta$ (J2000)} &
\colhead{$I$} &
\colhead{$(R-I)$} &
\colhead{$v$} &
\colhead{$v_{\rm err}$} &
\colhead{S/N} &
\colhead{Mask} \\
\colhead{}&
\colhead{(h$\,$ $\,$ m$\,$ $\,$s)} &
\colhead{($^\circ\,$ $\,'\,$ $\,''$)} &
\colhead{(mag)} &
\colhead{(mag)} &
\colhead{(\kms)} &
\colhead{(\kms)} &
\colhead{ } &
\colhead{ } 
}
\startdata
        2186 &  00 39 15.0  & +41 53 19.8 &    20.7 &    1.03 & $-$195.9 &    11.5 &   15.7 & N205-2 \\
        2196 &  00 39 15.9  & +41 52 23.0 &    20.7 &    0.89 & $-$309.8 &    11.5 &   11.7 & N205-2 \\
        2272 &  00 39 23.6  & +41 50 29.5 &    20.7 &    0.93 & $-$213.8 &    11.5 &   13.6 & N205-2 \\
        2441 &  00 39 25.3  & +41 48 39.8 &    20.9 &    0.84 & $-$240.8 &    11.5 &    9.0 & N205-2 \\
        2658 &  00 39 20.3  & +41 52 21.7 &    20.9 &    1.30 & $-$162.9 &    11.5 &   12.6 & N205-2 \\
        2681 &  00 39 18.5  & +41 54 13.7 &    21.0 &    0.91 & $-$261.8 &    11.5 &   10.5 & N205-2 \\
        2770 &  00 39 15.8  & +41 51 23.7 &    21.0 &    1.07 & $-$225.8 &    11.5 &    9.8 & N205-2 \\
        2805 &  00 39 20.9  & +41 53 40.0 &    21.1 &    0.99 & $-$291.8 &    11.5 &   10.1 & N205-2 \\
        3047 &  00 39 25.1  & +41 52 54.9 &    21.2 &    0.77 & $-$174.9 &    11.5 &    7.7 & N205-2 \\
        3234 &  00 39 22.5  & +41 53 31.4 &    21.4 &    0.69 &  $-$96.9 &    11.5 &    7.1 & N205-2 \\
        3238 &  00 39 23.7  & +41 51 54.1 &    21.4 &    0.66 & $-$288.8 &    11.5 &    7.3 & N205-2 \\
        3239 &  00 39 18.3  & +41 54 02.4 &    21.3 &    0.79 & $-$171.9 &    11.5 &    8.7 & N205-2 \\
        3279 &  00 39 19.3  & +41 52 28.5 &    21.3 &    0.90 & $-$165.9 &    11.5 &    7.9 & N205-2 \\
        3804 &  00 39 25.5  & +41 54 17.3 &    21.6 &    0.86 & $-$318.8 &    11.5 &    7.0 & N205-2 \\
        4716 &  00 39 30.2  & +41 50 03.2 &    20.1 &    1.87 & $-$393.7 &    11.5 &   30.8 & N205-2 \\
        4773 &  00 39 57.6  & +41 43 02.7 &    20.6 &    0.29 & $-$219.8 &    11.5 &   18.6 & N205-2 \\
        4832 &  00 40 02.2  & +41 42 56.5 &    20.8 &    0.64 & $-$168.9 &    11.5 &   21.9 & N205-2 \\
        4914 &  00 39 50.4  & +41 49 56.8 &    20.5 &    0.71 & $-$201.9 &    11.5 &   18.9 & N205-2 \\
        5002 &  00 39 36.5  & +41 48 13.7 &    20.6 &    0.78 & $-$516.6 &    11.5 &   16.3 & N205-2 \\
        5096 &  00 40 02.9  & +41 45 56.0 &    20.5 &    1.30 & $-$165.9 &    11.5 &   17.8 & N205-2 \\
        5100 &  00 39 58.3  & +41 48 33.2 &    20.5 &    1.35 & $-$162.9 &    11.5 &   18.0 & N205-2 \\
        5109 &  00 39 52.1  & +41 49 38.8 &    20.5 &    1.01 & $-$267.8 &    11.5 &   18.9 & N205-2 \\
        5123 &  00 39 57.2  & +41 48 09.3 &    20.6 &    0.97 & $-$234.8 &    11.5 &   17.9 & N205-2 \\
        ... &  ... & ... &   ... &    ... & ... &    ...&  ... & ... \\
\enddata
\tablecomments{Velocity measurements for individual stars in NGC~205.
Star names, positions and magnitudes are taken from \citet{dem03}.  We
list the heliocentric radial velocity ($v$), velocity error ($v_{\rm
err}$), the median per pixel signal-to-noise (S/N) and the DEIMOS mask name for each star.
Velocity error bars were determined from measurement overlaps for each
mask as discussed in \S\,\ref{subsec_rvel}.  We also listed the
median per pixel S/N value for each slitlet in the wavelength range
$8000$ to $9000\mbox{\AA}$ and refer the reader to Gilbert et
al.~(2005) for a discussion of velocity errors for individual slitlets
based on the S/N.  Table~4 is published in its entirety in the
electronic edition of the {\it Astronomical Journal} and is available upon
request.  A portion is
shown here for guidance regarding its form and content.}
\end{deluxetable}

\begin{deluxetable}{cccrcccc}
\tabletypesize{\scriptsize}
\tablecaption{Major-Axis Velocity and Velocity Dispersion Profile of NGC~205}
\tablewidth{0pt}
\tablehead{
\colhead{Semi-major Dist.} &
\colhead{$\alpha$ (J2000)} &
\colhead{$\delta$ (J2000)} &
\colhead{$v_{\rm rot}$} &
\colhead{$v_{\rm err}$} &
\colhead{$\sigma$} &
\colhead{$\sigma_{\rm err}$} \\
\colhead{(arcmin)}&
\colhead{(h$\,$ $\,$ m$\,$ $\,$s)} &
\colhead{($^\circ\,$ $\,'\,$ $\,''$)} &
\colhead{(\kms)} &
\colhead{(\kms)} &
\colhead{(\kms)} &
\colhead{(\kms)} }
\startdata
 $-$17.3 &  00 41 12.4  & +41 29 19.7 & $-$299.1 &     8.9 &  & \\
 $-$14.0 &  00 41 01.5  & +41 31 25.3 & $-$292.0 &    10.0 &     34.4 &     9.4  \\
 $-$10.8 &  00 40 53.5  & +41 33 42.1 & $-$267.1 &     9.2 &  & \\
  $-$8.4 &  00 40 42.3  & +41 34 46.9 & $-$258.5 &     4.7 &     35.6 &     7.6  \\
  $-$7.0 &  00 40 32.3  & +41 34 46.2 & $-$247.2 &     6.7 &  & \\
  $-$6.0 &  00 40 25.9  & +41 35 08.5 & $-$241.4 &     8.5 &     44.9 &     7.4  \\
  $-$5.0 &  00 40 23.9  & +41 36 09.0 & $-$232.6 &     5.3 &  & \\
  $-$4.0 &  00 40 22.1  & +41 37 07.0 & $-$239.6 &     5.5 &     30.0 &     5.0  \\
  $-$3.0 &  00 40 23.6  & +41 38 11.8 & $-$239.0 &     3.9 &  & \\
  $-$2.0 &  00 40 24.4  & +41 39 12.6 & $-$239.4 &     3.9 &     27.2 &     4.9  \\
  $-$1.0 &  00 40 28.4  & +41 40 20.3 & $-$243.0 &     5.3 &  & \\
    0.0 &  00 40 30.4  & +41 41 22.9 & $-$246.9 &     5.7 &     22.4 &     6.0  \\
    1.0 &  00 40 26.3  & +41 42 15.8 & $-$249.9 &     3.8 &  & \\
    2.0 &  00 40 21.0  & +41 43 05.9 & $-$253.5 &     3.1 &     35.7 &     4.9  \\
    3.0 &  00 40 18.9  & +41 44 04.2 & $-$254.1 &     4.5 &  & \\
    4.0 &  00 40 23.8  & +41 45 07.2 & $-$253.2 &     3.8 &     37.3 &     6.1  \\
    5.0 &  00 40 22.3  & +41 46 12.0 & $-$251.1 &     3.8 &  & \\
    6.0 &  00 40 24.1  & +41 47 38.8 & $-$240.9 &     5.7 &     45.8 &     6.7  \\
    7.0 &  00 40 15.1  & +41 47 57.1 & $-$229.9 &     5.9 &  & \\
    8.8 &  00 39 58.0  & +41 47 13.2 & $-$222.7 &     8.3 &     40.1 &     8.3  \\
   11.5 &  00 39 48.7  & +41 48 43.2 & $-$218.4 &     6.2 &  & \\
   14.1 &  00 39 36.2  & +41 49 45.8 & $-$225.2 &     8.1 &     38.4 &     8.0  \\
\enddata

\tablecomments{The major-axis velocity and velocity dispersion profile
determined from the combined measurements of individual RGB stars.
Positive radial bins correspond to the NW side of NGC~205, on the side
of NGC~205 farthest from M31.  The profiles are determined
along the major-axis of NGC~205 which forms a gentle 'S' curve as
shown in Figure~\ref{fig_imvel}.}
\end{deluxetable}

\clearpage

\vskip 8cm

\begin{thebibliography}{47}
\expandafter\ifx\csname natexlab\endcsname\relax\def\natexlab#1{#1}\fi

\bibitem[{{Beasley} {et~al.}(2005){Beasley}, {Strader}, {Brodie}, {Cenarro}, \&
  {Geha}}]{bea05}
{Beasley}, M., {Strader}, J., {Brodie}, J., {Cenarro}, J., \& {Geha}, M. 2005,
  \apj, submitted

\bibitem[{{Bender} \& {Nieto}(1990)}]{ben90}
{Bender}, R., \& {Nieto}, J.-L. 1990, \aap, 239, 97

\bibitem[{{Bender} {et~al.}(1991){Bender}, {Paquet}, \& {Nieto}}]{ben91}
{Bender}, R., {Paquet}, A., \& {Nieto}, J.-L. 1991, \aap, 246, 349

\bibitem[{{Binggeli} {et~al.}(1988){Binggeli}, {Sandage}, \& {Tammann}}]{bin88}
{Binggeli}, B., {Sandage}, A., \& {Tammann}, G.~A. 1988, \araa, 26, 509

\bibitem[{{Binney} \& {Tremaine}(1987)}]{bin87}
{Binney}, J., \& {Tremaine}, S. 1987, {Galactic Dynamics} (Princeton, NJ,
  Princeton University Press, 1987)

\bibitem[{{Bullock} {et~al.}(2000){Bullock}, {Kravtsov}, \& {Weinberg}}]{bul00}
{Bullock}, J.~S., {Kravtsov}, A.~V., \& {Weinberg}, D.~H. 2000, \apj, 539, 517

\bibitem[{{Cappellari} {et~al.}(1999){Cappellari}, {Bertola}, {Burstein},
  {Buson}, {Greggio}, \& {Renzini}}]{cap99}
{Cappellari}, M., {Bertola}, F., {Burstein}, D., {Buson}, L.~M., {Greggio}, L.,
  \& {Renzini}, A. 1999, \apjl, 515, L17

\bibitem[{{Carter} \& {Sadler}(1990)}]{car90}
{Carter}, D., \& {Sadler}, E.~M. 1990, \mnras, 245, 12P

\bibitem[{{Choi} {et~al.}(2002){Choi}, {Guhathakurta}, \& {Johnston}}]{cho02}
{Choi}, P.~I., {Guhathakurta}, P., \& {Johnston}, K.~V. 2002, \aj, 124, 310

\bibitem[{{Cooper et al.~}(2006)}]{coo06}
{Cooper et al.~}. 2006, in prep

\bibitem[{{Corradi} {et~al.}(2005){Corradi}, {Magrini}, {Greimel}, {Irwin},
  {Leisy}, {Lennon}, {Mampaso}, {Perinotto}, {Pollacco}, {Walsh}, {Walton}, \&
  {Zijlstra}}]{cor05}
{Corradi}, R.~L.~M., {Magrini}, L., {Greimel}, R., {Irwin}, M., {Leisy}, P.,
  {Lennon}, D.~J., {Mampaso}, A., {Perinotto}, M., {Pollacco}, D.~L., {Walsh},
  J.~R., {Walton}, N.~A., \& {Zijlstra}, A.~A. 2005, \aap, 431, 555

\bibitem[{{De Rijcke} {et~al.}(2001){De Rijcke}, {Dejonghe}, {Zeilinger}, \&
  {Hau}}]{der01}
{De Rijcke}, S., {Dejonghe}, H., {Zeilinger}, W.~W., \& {Hau}, G.~K.~T. 2001,
  \apjl, 559, L21

\bibitem[{{Dekel} \& {Silk}(1986)}]{dek86}
{Dekel}, A., \& {Silk}, J. 1986, \apj, 303, 39

\bibitem[{{Demers} {et~al.}(2003){Demers}, {Battinelli}, \& {Letarte}}]{dem03}
{Demers}, S., {Battinelli}, P., \& {Letarte}, B. 2003, \aj, 125, 3037

\bibitem[{{Faber} {et~al.}(2003){Faber}, {Phillips}, {Kibrick}, {Alcott},
  {Allen}, {Burrous}, {Cantrall}, {Clarke}, {Coil}, {Cowley}, {Davis}, {Deich},
  {Dietsch}, {Gilmore}, {Harper}, {Hilyard}, {Lewis}, {McVeigh}, {Newman},
  {Osborne}, {Schiavon}, {Stover}, {Tucker}, {Wallace}, {Wei}, {Wirth}, \&
  {Wright}}]{fab03}
  {Faber}, S.~M., et~al.~2003, in Instrument Design and
  Performance for Optical/Infrared Ground-based Telescopes. Edited by Iye,
  Masanori; Moorwood, Alan F. M. Proceedings of the SPIE, Volume 4841.,
  1657--1669

\bibitem[{{Ferguson} {et~al.}(2002){Ferguson}, {Irwin}, {Ibata}, {Lewis}, \&
  {Tanvir}}]{fer02}
{Ferguson}, A.~M.~N., {Irwin}, M.~J., {Ibata}, R.~A., {Lewis}, G.~F., \&
  {Tanvir}, N.~R. 2002, \aj, 124, 1452

\bibitem[{Font {et~al.}(2005)Font, Johnston, Bullock, \& Robertson}]{fon05}
Font, A.~S., Johnston, K.~V., Bullock, J.~S., \& Robertson, B. 2005

\bibitem[{{Freedman} {et~al.}(2001){Freedman}, {Madore}, {Gibson}, {Ferrarese},
  {Kelson}, {Sakai}, {Mould}, {Kennicutt}, {Ford}, {Graham}, {Huchra},
  {Hughes}, {Illingworth}, {Macri}, \& {Stetson}}]{fre01}
{Freedman}, W.~L., {Madore}, B.~F., {Gibson}, B.~K., {Ferrarese}, L., {Kelson},
  D.~D., {Sakai}, S., {Mould}, J.~R., {Kennicutt}, R.~C., {Ford}, H.~C.,
  {Graham}, J.~A., {Huchra}, J.~P., {Hughes}, S.~M.~G., {Illingworth}, G.~D.,
  {Macri}, L.~M., \& {Stetson}, P.~B. 2001, \apj, 553, 47

\bibitem[{{Geehan} {et~al.}(2005){Geehan}, {Fardal}, {Babul}, \&
  {Guhathakurta}}]{gee05}
{Geehan}, J., {Fardal}, M., {Babul}, A., \& {Guhathakurta}, P. 2005,
  Astro-ph/0501241

\bibitem[{{Geha} {et~al.}(2002){Geha}, {Guhathakurta}, \& {van der
  Marel}}]{geh02}
{Geha}, M., {Guhathakurta}, P., \& {van der Marel}, R.~P. 2002, \aj, 124, 3073

\bibitem[{{Geha} {et~al.}(2003){Geha}, {Guhathakurta}, \& {van der
  Marel}}]{geh03}
---. 2003, \aj, 126, 1794

\bibitem[{{Guhathakurta} {et~al.}(2004){Guhathakurta}, {Rich}, {Reitzel},
  {Cooper}, {Gilbert}, {Majewski}, {Ostheimer}, {Geha}, K., \&
  {Patterson}}]{guh04}
{Guhathakurta}, P., {Rich}, M., {Reitzel}, D., {Cooper}, M., {Gilbert}, K.,
  {Majewski}, S., {Ostheimer}, J., {Geha}, M., K., J., \& {Patterson}, R. 2004,
  Astro-ph/0406145

\bibitem[{{Haas}(1998)}]{haa98}
{Haas}, M. 1998, \aap, 337, L1

\bibitem[{{Hodge}(1973)}]{hod73}
{Hodge}, P.~W. 1973, \apj, 182, 671

\bibitem[{{Horne}(1986)}]{hor86}
{Horne}, K. 1986, \pasp, 98, 609

\bibitem[{{Ibata} {et~al.}(2004){Ibata}, {Chapman}, {Ferguson}, {Irwin},
  {Lewis}, \& {McConnachie}}]{iba04}
{Ibata}, R., {Chapman}, S., {Ferguson}, A.~M.~N., {Irwin}, M., {Lewis}, G., \&
  {McConnachie}, A. 2004, \mnras, 351, 117

\bibitem[{{Ibata} {et~al.}(2001){Ibata}, {Irwin}, {Lewis}, {Ferguson}, \&
  {Tanvir}}]{iba01}
{Ibata}, R., {Irwin}, M., {Lewis}, G., {Ferguson}, A.~M.~N., \& {Tanvir}, N.
  2001, \nat, 412, 49

\bibitem[{{Ibata} {et~al.}(1995){Ibata}, {Gilmore}, \& {Irwin}}]{iba95}
{Ibata}, R.~A., {Gilmore}, G., \& {Irwin}, M.~J. 1995, \mnras, 277, 781

\bibitem[{{Kent}(1987)}]{ken87}
{Kent}, S.~M. 1987, \aj, 94, 306

\bibitem[{{Kleyna} {et~al.}(2003){Kleyna}, {Wilkinson}, {Gilmore}, \&
  {Evans}}]{kle03}
{Kleyna}, J.~T., {Wilkinson}, M.~I., {Gilmore}, G., \& {Evans}, N.~W. 2003,
  \apjl, 588, L21

\bibitem[{{Lee}(1996)}]{lee96}
{Lee}, M.~G. 1996, \aj, 112, 1438

\bibitem[{Mastropietro {et~al.}(2004)}]{mas04}
Mastropietro, C., {et~al.} 2004

\bibitem[{{Mayer} {et~al.}(2001){Mayer}, {Governato}, {Colpi}, {Moore},
  {Quinn}, {Wadsley}, {Stadel}, \& {Lake}}]{may01}
{Mayer}, L., {Governato}, F., {Colpi}, M., {Moore}, B., {Quinn}, T., {Wadsley},
  J., {Stadel}, J., \& {Lake}, G. 2001, \apj, 559, 754

\bibitem[{{McConnachie} {et~al.}(2005){McConnachie}, {Irwin}, {Ferguson},
  {Ibata}, {Lewis}, \& {Tanvir}}]{mcc05}
{McConnachie}, A.~W., {Irwin}, M.~J., {Ferguson}, A.~M.~N., {Ibata}, R.~A.,
  {Lewis}, G.~F., \& {Tanvir}, N. 2005, \mnras, 356, 979

\bibitem[{{McConnachie} {et~al.}(2004){McConnachie}, {Irwin}, {Lewis}, {Ibata},
  {Chapman}, {Ferguson}, \& {Tanvir}}]{mcc04}
{McConnachie}, A.~W., {Irwin}, M.~J., {Lewis}, G.~F., {Ibata}, R.~A.,
  {Chapman}, S.~C., {Ferguson}, A.~M.~N., \& {Tanvir}, N.~R. 2004, \mnras, 351,
  L94

\bibitem[{{Meylan} {et~al.}(2001){Meylan}, {Sarajedini}, {Jablonka},
  {Djorgovski}, {Bridges}, \& {Rich}}]{mey01}
{Meylan}, G., {Sarajedini}, A., {Jablonka}, P., {Djorgovski}, S.~G., {Bridges},
  T., \& {Rich}, R.~M. 2001, \aj, 122, 830

\bibitem[{{Moore} {et~al.}(1998){Moore}, {Lake}, \& {Katz}}]{moo98}
{Moore}, B., {Lake}, G., \& {Katz}, N. 1998, \apj, 495, 139

\bibitem[{{Peletier}(1993)}]{pel93}
{Peletier}, R.~F. 1993, \aap, 271, 51

\bibitem[{{Peterson} \& {Caldwell}(1993)}]{pet93}
{Peterson}, R.~C., \& {Caldwell}, N. 1993, \aj, 105, 1411

\bibitem[{{Reitzel} {et~al.}(2004){Reitzel}, {Guhathakurta}, \& {Rich}}]{rei04}
{Reitzel}, D., {Guhathakurta}, P., \& {Rich}, R. 2004, \aj, 127, 2133

\bibitem[{{Schlegel} {et~al.}(1998){Schlegel}, {Finkbeiner}, \&
  {Davis}}]{sch98}
{Schlegel}, D.~J., {Finkbeiner}, D.~P., \& {Davis}, M. 1998, \apj, 500, 525

\bibitem[{{Simien} \& {Prugniel}(2002)}]{sim02}
{Simien}, F., \& {Prugniel}, P. 2002, \aap, 384, 371

\bibitem[{{Thilker} {et~al.}(2004){Thilker}, {Braun}, {Walterbos}, {Corbelli},
  {Lockman}, {Murphy}, \& {Maddalena}}]{thi04}
{Thilker}, D.~A., {Braun}, R., {Walterbos}, R.~A.~M., {Corbelli}, E.,
  {Lockman}, F.~J., {Murphy}, E., \& {Maddalena}, R. 2004, \apjl, 601, L39

\bibitem[{{Toomre} \& {Toomre}(1972)}]{too72}
{Toomre}, A., \& {Toomre}, J. 1972, \apj, 178, 623

\bibitem[{{van Zee} {et~al.}(2004){van Zee}, {Skillman}, \& {Haynes}}]{van04}
{van Zee}, L., {Skillman}, E.~D., \& {Haynes}, M.~P. 2004, \aj, 128, 121

\bibitem[{{Walterbos} \& {Kennicutt}(1988)}]{wal88}
{Walterbos}, R.~A.~M., \& {Kennicutt}, R.~C. 1988, \aap, 198, 61

\bibitem[{{Welch} {et~al.}(1998){Welch}, {Sage}, \& {Mitchell}}]{wel98}
{Welch}, G.~A., {Sage}, L.~J., \& {Mitchell}, G.~F. 1998, \apj, 499, 209

\bibitem[{{Young} \& {Lo}(1997)}]{you97}
{Young}, L.~M., \& {Lo}, K.~Y. 1997, \apj, 476, 127

\end{thebibliography}
\end{document}